# Finite size correction to the thermo-magnetic properties of charged Bose gas

A. H. Abbas*, W. M. Seif*, Th. elsherbini*, Ahmed S. Hassan**

*physics department, faculty of science, Cairo university, Egypt.

**physics department, faculty of science, Menia university, Egypt.

Abstract

We consider the finite size effect on thermo-magnetic properties of charged Bose gases confined in a quasi two-dimentsional potential. A modified semiclassical approach is used, taking into account the finite size correction to guarantee an accurate density of states. The charged spinless Bose gas shows diamagnetic behavior. Temperature dependent magnetization is indicated for the investigated system that condense at a resistive critical temperature in accordance with the applied magnetic field.

## 1 Introduction

Bosons with charge and spin degrees of freedom can manifest diamagnetic or paramagnetic properties [1]. The paramagnetic behavior, or the so-called *Pauli paramagnetism*, is caused by the orientation of spin dipole moments along the field direction. The diamagnetic behavior (*Landau diamagnetism*) is due to the induction of orbital motion in the material which results in orbital dipole moments oppose those of the external field, according to *Lenz rule*. The induced field in diamagnetic material, in extreme cases, may completely compensate the external field and then the interior of the material becomes absolutely free of fields. This phenomenon observed in superconductivity is known as *Meissner effect*. Therefore, paramagnetic materials enforce the external magnetic field while diamagnetic materials hinder it. However, the system exhibits diamagnetic behavior when the magnetization is negative ($M < 0$) and it tends to be paramagnetic if the magnetization is positive ($M > 0$). This can be understood by calculating the magnetization as a function of magnetic field $M(B)$.

Here, we focus on the magnetic properties of charged spinless Bose gases. Charged Bose gas possesses interesting magnetic properties owing to the quantization of the orbital motion of charged particles in a magnetic field. It is supposed to exhibit *landau diamagnetic* behavior due to the charge degree of freedom. Numerous studies have been devoted to investigate the properties of CBG and its behavior. The simultaneous studies of Schafroth and Blatt *et al*. pointed out that an ideal CBG could show essential equilibrium features of superconductor [2] and superfluid [5]. It was found that even a small value of magnetic field can eliminate BEC in 3-dimensions. Below the transition temperature, the number of superconducting concurrence bosons was indicated to be temperature dependent [2]. Extending the space dimensionality of the system to D>3, May [6] found mathematically that CBG might condense in a homogeneous magnetic field only when D>4, which is rather unphysical condition. The diamagnetic property for CBG has been used to explain the normal-state diamagnetism above the critical energy (Tc) of superconductors [9]. The normal-state diamagnetism appearing in many cuprates [8] was mentioned as one of the evidences for charged real-space bosons. This may shed light on some of the properties of CBGs.

In the present study, we aim to carry out a detailed investigation of the finite size effect on a system of 3D spinless charged Bose gas. We consider CBG that is confined in *xy* plane superimposed with magnetic field pointed along the *z* direction. The thermodynamic properties depend on the construction of the density of states (DOS). Therefore, DOS including the finite size effect is calculated. This argument meets reality. However, the real condensate cloud consists of finite number of atoms. The outline of this paper is as follows. In the next section, the Hamiltonian of the system is written down and the modified DOS is then derived. The critical temperature and the magnetization for CBG are given in Sec. 3. In Sec. 4, we discussed the obtained results in details. Finally, we summarized the conclusions in the last section.

## 2    The physical model

Let us consider a system composed of Bose gas with mass *m* and charge *q* which is confined in two-dimensional (2D) harmonic potential. The system is exposed to a uniform magnetic field aligned along the *z*-axis, $\hat{B}$ = $B\hat{e}_z$. The Hamiltonian can be expressed as

$$H = \frac{1}{2m}(p - Aq)^2 + \frac{1}{2}m\omega_\perp(x^2 + y^2) \quad (1)$$

where $\omega_\perp = \omega_{xy}$ is the effective trapping frequency. Choosing the gauge potential $A = \frac{1}{2}B \times r$

$$H = H_r + H_z$$

where

$$H_z = \frac{p_z^2}{2m}$$

is associated with the translational degree of freedom of the system in the *z*-direction, with effective mass *m*. Here, the Hamiltonian

$$H_r = \frac{p_x^2 + p_y^2}{2m} + \frac{1}{2}m(\omega_\perp^2 + \Omega^2)(x^2 + y^2) - \Omega L_z \quad (2)$$

Describe the behavior in xy plane. It shows that the system is trapped with frequency ($\omega_\perp^2 + \Omega^2$) with $qB = \frac{\Omega}{2m}$ is the cyclotron frequency. The eigenvalues of the Hamiltonian in equation (2) are given by [10]

$$E_{n_+,n_-,k_z} = \frac{\hbar^2 k_z^2}{2m} + n_+\hbar\omega_+ + n_-\hbar\omega_- + E_0 \quad (3)$$

Where $\omega_\pm = \omega \mp \Omega$, $E_0 = \hbar(\omega_\perp^2 + \Omega^2)^{1/2}$ and $n_\pm$ is non-negative integers. Clearly, the ground state for charged Bose gas has a magnetic property. the energy eigenvalues show that the system has translational degree of freedom in direction of the applied field and hyperfine vibrational degrees of freedom labelled by *n+* in addition to degenerate Landau level indexed by *n–*.

## 1    Accurate DOS.

The accurate density of states for the system can be calculated using the method that was used by Kirsten and Toms [11], in which an asymptotic high- temperature expansion of the partition function $Z(\beta)$ is obtained. Generally, consider the eigenvalues $E_{k_z,n_+,n_-}$ (8), we have,

$$\rho_s(E) = \frac{4}{c_1} E^{3/2} + c_2 E^{1/2}$$

Where

$$c_1 = (\frac{m}{2\pi^2\hbar^2})^{\frac{1}{2}} \frac{1}{\hbar\omega_+\omega_-}$$
$$c_2 = c_1 \hbar(\omega_+ + \omega_-)$$

The q-potential for Bose gas using the eigenvalues [8] given by,

$$q = - \sum_{n_+,n_-,k_z} \ln(1 - e^{E_{n_+,n_-,k_z}})$$

Using the semiclassical approximation (in which the summation over n is converted into an integral weighted by an accurate DOS) to calculate q requires expanding the logarithm to express the q potential as a sum over Bose-Einstein distribution:

$$q = q_0 + \int \rho(E) e^{-j\beta E_{n_+,n_-,k_z}} dE \tag{4}$$

where $q_0 = -\ln(1-z)$ is the contribution of the ground states, $z = e^{-\beta(E_0-\mu)}$ is the effective fugacity and $\beta = 1/kT$. Using Dos [12] leads to, where $g_\nu(z) = \sum \frac{z^j}{j^\nu}$ is the usual Bose function, reduces to Riemann zeta function $\zeta(n)$ at $z = 1$. Now, it is straightforward to calculate the thermodynamical properties of the system.

The thermodynamic potential (7) now reads

$$q = q_0 + \frac{l_z}{(2\pi)^{\frac{1}{2}}l_\perp}(\frac{1}{\hbar\omega_\perp})^{\frac{5}{2}}((k_B T)^{\frac{5}{2}} g_{\frac{7}{2}}(z) + \hbar\omega_\perp(1+\overline{B})(k_B T)^{\frac{3}{2}} g_{\frac{5}{2}}(z)) \tag{5}$$

Where $\overline{B} = \frac{\Omega}{\omega_\perp}$ and $l_\perp = (\frac{m}{\hbar\omega_\perp})^{\frac{1}{2}}$

## 2.1 Condensation fraction and critical temperature

In terms of the q-potential, the total number of particles is derived as $N = z\frac{\partial q}{\partial z}$

$$N = N_0 + \frac{l_z}{(2\pi)^{\frac{1}{2}}l_\perp}(\frac{1}{\hbar\omega_\perp})^{\frac{5}{2}}((k_B T)^{\frac{5}{2}} g_{\frac{5}{2}}(z) + \hbar\omega_\perp(1+\overline{B}^2)(k_B T)^{\frac{3}{2}} g_{\frac{3}{2}}(z)) \tag{6}$$

$N_0$ is the number of particle condensate in the ground state of the system. In thermodynamic limit, the BEC transition temperature for the system,

$$k_B T_0 = \hbar\omega_\perp (\frac{N}{\zeta(5/2)})^{2/5} (\frac{(2\pi)^{\frac{1}{2}}l_\perp}{l_z})^{2/5} \tag{7}$$

it just for the critical BEC temperature for ideal gas of $N$ particles confined in trap with the frequency $\omega_\perp$ in $x$, $y$ plane and free in $z-$ direction.
The condensate fraction in terms of the reduced temperature $\tau = T/T_0$ is given by

$$\frac{N_0}{N} = 1 - \tau^{\frac{5}{2}} + R(\underline{B})\tau^{\frac{3}{2}} \tag{8}$$

where

$$R(\overline{B}) = (1 + \overline{B}^2) \left(\frac{\zeta(5/2)}{N}\right)^{2/5} \left(\frac{l_z}{(2\pi)^{\frac{1}{2}} l_\perp}\right)^{2/5} \frac{\zeta(3/2)}{\zeta(5/2)} \tag{9}$$

The parameter $R(\overline{B})$ amounts to the finite size effect. The condensation fraction suppressed as the magnetic field is strengthened. Its consistent with our intuition that small value of the magnetic field can eliminate BEC in a charged Bose gas [5]. Considering the thermodynamic limit $N \to \infty$ the ideal behavior is recovered. It is important to consider the effect of the finite size on the transition temperature, which can be calculated by setting $N/N_o$ in Eq. (19) equal to zero, thus

$$T(\overline{B}) = T_0\left(1 - \frac{2}{5} R(\overline{B})\right) \tag{10}$$

the system condenses at temperature relevant to applied field. The transition temperature decreases as the magnetic field is strengthened which agree with our intuition established in the study of superconductivity. In thermodynamic limit and absence of magnetic field the critical temperature is the same as the ideal gas critical temperature $T_o$.

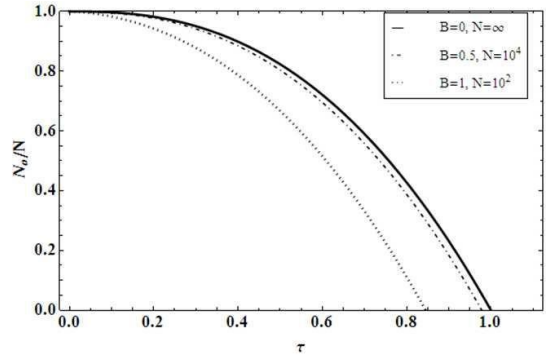

Figure 1: Condensation fraction versus reduced temperature

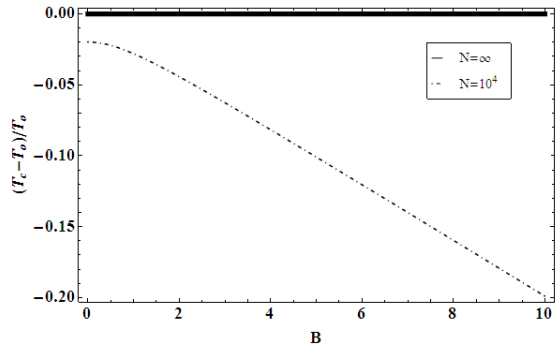

Figure 2: fraction of reduced temperature versus applied field

## 2.2  Magnetization and magnetic susceptibility

Magnetization at temperature above BEC transition temperature for ideal gas $T_o$ is given by,

$$M = -(M_0 - \frac{l_z}{(2\pi)^{\frac{1}{2}}l_\perp}(\frac{1}{\hbar\omega_\perp})^{\frac{5}{2}}\frac{\hbar\omega_\perp B}{(1+\underline{B}^2)^{\frac{1}{2}}}(k_BT)^{\frac{5}{2}}g_{\frac{5}{2}}(z)) \qquad (11)$$

Where $M_0$ is the magnetization due to condensate part. The total magnetization per particle in terms of the reduced temperature now reads

$$M = -(M_0 - \frac{\hbar\omega_\perp B}{(1+\underline{B}^2)^{\frac{1}{2}}}\tau^{\frac{5}{2}}g_{\frac{5}{2}}(z)/\zeta(5/2)) \qquad (12)$$

For $T \leq T_o$

$$M = -(M_0 - \frac{\hbar\omega_\perp B}{(1+\underline{B}^2)^{\frac{1}{2}}}\tau^{\frac{5}{2}}) \qquad (13)$$

Due to charge degree of freedom CBG exhibits strong landau diamagnetism.

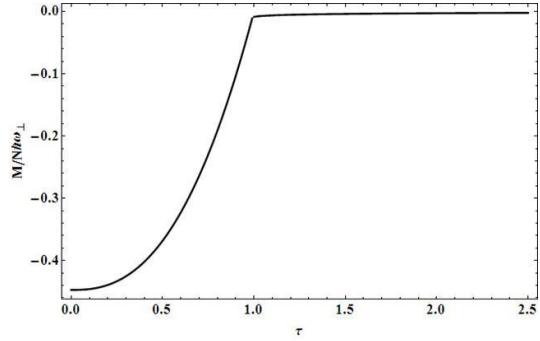

Figure 3: Magnetization versus reduced temperature

Diamagnetic behavior is due to the induction of orbital motion in the material (Landau diamagnetism), results in orbital dipole moments oppose those of the external field according to Lenz rule. The induced field in diamagnetic material in extreme case may completely compensate the external field and the interior of the material is absolutely free of fields. This phenomenon observed in superconductivity. The present result shows that $M$ vanishes as $B \to 0$ at all temperatures, implying that the Meissner Ochsenfeld effect might not exist in a trapped CBG. Above the critical temperature, the magnetization vanishes with temperature and the magnetic field penetrates the system. It's important to investigate the magnetic susceptibility to address the nature of magnetic transition. The magnetic susceptibility above the critical temperature is obtained as

$$\chi = -\chi_0(1 - \frac{g_{\frac{5}{2}}(z)}{\zeta(\frac{5}{2})}\tau^{\frac{5}{2}} - \underline{B}(1+\underline{B}^2)\frac{g_{\frac{3}{2}}(z)}{\zeta(\frac{5}{2})}\tau^{\frac{3}{2}}(\frac{1}{z}\frac{dz}{d\underline{B}})) \qquad (14)$$

Where $\chi_0 = \frac{\hbar\omega_\perp}{(1+\underline{B}^2)^{3/2}}$

$$\frac{1}{z}\frac{dz}{d\underline{B}} = -\frac{g_{\frac{5}{2}}(z)+\underline{B}(1+\underline{B}^2)^{1/2}\frac{g_{\frac{3}{2}}(z)}{\tau}}{g_{\frac{3}{2}}(z)+\underline{B}(1+\underline{B}^2)^{1/2}\frac{g_{\frac{1}{2}}(z)}{\tau}} \qquad (15)$$

whereas below the critical temperature

$$\chi = -\chi_0(1 - \tau^{\frac{5}{2}}) \qquad (16)$$

The system exhibit a strong diamagnetism below the transition temperature.

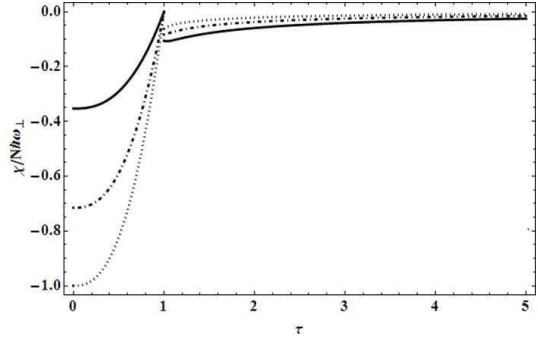

Figure 5: magnetic susceptibility versus reduced temperature

It has a negative value approximately constant at very low temperature. The temperature dependence of the magnetic susceptibility appears as temperature rises; it reaches to zero at $T_c$ at which magnetic field penetrates the system. For sufficient strong value of the external magnetic field, the system displays a diamagnetic behavior above $T_c$ with negative susceptibility. Again the susceptibility goes to zero at relatively high temperature.

# 6  Conclusion

We have investigated systems of charged bosons in a magnetic field. Considering accurate density of states, both systems exhibit temperature dependent magnetization. They condense at a critical temperature relevant to the applied magnetic field. The charged spinless Bose gas exhibits diamagnetic properties. This is due to the orbital moment associated with the motion of charges which opposes the external field. In accordance with the results obtained in Ref. [9] in the thermodynamic limit N→∞.